\documentclass[pra,twocolumn,showpacs,nofootinbib,floatfix]{revtex4}

\pdfoutput=1

\usepackage{amsmath,amsfonts,amssymb,bm}
\usepackage{dcolumn}
\usepackage[final]{graphicx}

\begin{document}

\bibliographystyle{apsrev}

\title{Interaction of single molecules with metallic nanoparticles}

\author{Ulrich Hohenester}\email{ulrich.hohenester@uni-graz.at}
\author{Andreas Tr\"ugler}
\affiliation{Institut f\"ur Physik,
  Karl--Franzens--Universit\"at Graz, Universit\"atsplatz 5,
  8010 Graz, Austria}

\date{\today}

\begin{abstract}

We theoretically investigate the interaction between a single molecule and a metallic nanoparticle. We develop a general quantum mechanical description for the calculation of the enhancement of radiative and non-radiative decay channels for a molecule situated in the nearfield regime of the metallic nanoparticle. Using a boundary element method approach, we compute the scattering rates for several nanoparticle shapes. We also introduce an eigenmode expansion and quantization scheme for the surface plasmons, which allows us to analyze the scattering processes in simple physical terms. An intuitive explanation is given for the large quantum yield of quasi one- and two-dimensional nanostructures. Finally, we briefly discuss resonant F\"orster energy transfer in presence of metallic nanoparticles.

\end{abstract}



\maketitle


\section{Introduction}

Metal nanoparticles can sustain local surface plasmon excitations, particle plasmons, which are hybrid modes of a light field coupled to a coherent electron charge oscillation~\cite{kreibig:95,novotny:06}. The properties of these excitations depend strongly on particle geometry and interparticle coupling, and give rise to a variety of effects, such as frequency-dependent absorption and scattering or near field enhancement. Particle plasmons enable the concentration of light fields to nanoscale volumes and play a key role in surface enhanced spectroscopy~\cite{moskovits:85}. This can be exploited for a variety of applications, such as surface-enhanced Raman scattering \cite{chance:78,ford:84} or biochemical sensorics \cite{haes:02}. 

Improved nanofabrication methods nowadays allow advanced control of the nanoparticle shape and the arrangement of nanoparticle ensembles \cite{rechberger:03}, and open the possibility to flexibly tailor specific molecule-nanoparticle couplings. Anger et al. \cite{anger:06} recently investigated the fluorescence of a single molecule and a single laser-irradiated spherical gold nanoparticle, and demonstrated by varying the molecule-nanoparticle distance the continuous transition from fluorescence enhancement to fluorescence quenching \cite{dulkeith:05}. This finding was supported by K\"uhn and coworkers~\cite{kuehn:06} using a similar setup, and stimulated a large number of related experiments \cite{gerber.prb:07,ritman:07}.

The theoretical description of the modified molecule dynamics in presence of metallic nanoparticles (MNPs) has been addressed by several authors in the past (see, e.g. \cite{chance:78,agarwal:84,girard:95,girard:05b} and references therein). More recently, the question of how to treat quantum electrodynamics in the presence of absorbing media has seen anewed interest \cite{henry:96,scheel:98,vogel:06,raabe:07}, motivated by related work on thermal nearfield radiation \cite{carminati:99,dewilde:06} and thermally induced scattering processes of atoms located in the vicinity of metallic or superconducting bodies \cite{henkel:99,rekdal:04,skagerstam:06,hohenester.pra:07b}. This work has identified the Green tensor as the central element to establish, in linear response, a convenient link between quantum electrodynamics and classical Maxwell theory.

In this work we start by reviewing the description scheme for a molecular dipole placed in the vicinity of a metallic nanoparticle. We show how to use the dyadic Green function of Maxwell's theory to express the electric field fluctuations in terms of the current fluctuations in the metal, and how to derive the molecule scattering rate within linear response theory. In the quasistatic regime, where the light wavelength is much larger than the nanoparticle size and the molecule-MNP distance, the decay rate can be separated into a radiative and non-radiative contribution, which both show a simple scaling behavior. We next use a boundary element method approach \cite{hohenester.prb:05} to compute these decay rates for realistic nanoparticle geometries. The results are shown to be in nice agreement with Mie theory. We additionally introduce for a Drude description of the metal electrons a quantization scheme for the surface plasmons \cite{ritchie:57,barton:79}. Although this approach could be used to study the nonlinear optical response of metallic nanoparticles, in this work we use this framework to provide an intuitive picture for the radiative and non-radiative molecular decay channels only, and give a simple explanation for the higher quantum yield achievable in quasi one- and two-dimensional nanoparticles. Finally, we briefly discuss resonant F\"orster energy transfer between donor and acceptor molecules placed in the nanoparticle nearfield.


\section{Theory}

As for the description of the molecule, in this work we follow Refs.~\cite{girard:95,girard:05b} who considered a generic two-level system. This approach is also best suited for other quantum emitters, such as colloidal quantum dots. We describe the molecule in terms of a generic two level scheme, with $0$ the groundstate and $1$ the excited molecular state with energy $E_1$. Other, more refined description schemes, considering, e.g., the different vibronic states of the molecule \cite{xu:04,johansson:05} or details of the molecular orbitals \cite{andreussi:04}, could be treated in a similar fashion.

We first consider the optical decay from the excited state $1$ to the groundstate. Associated to this transition is a dipole operator
\begin{equation}\label{eq:dipole}
  \hat{\bm d}={\bm\mu}\left(\hat\sigma_- +\hat\sigma_+\right)\equiv
  \hat{\bm d}^++\hat{\bm d}^-\,,
\end{equation}
where $\bm\mu$ is the molecular dipole moment, assumed to be real, $\hat\sigma_+=|1\rangle\langle 0|$ is the molecular excitation operator which brings the molecule from the ground to the excited state, and $\hat\sigma_-=|0\rangle\langle 1|$ the corresponding de-excitation operator. In an interaction representation according to the free Hamiltonian $\hat H_0=E_1\,|1\rangle\langle 1|$ the operators $\hat\sigma_-$ and $\hat\sigma_+$ oscillate with positive and negative frequencies $e^{-i\omega t}$ and $e^{i\omega t}$, respectively, as indicated by the dipole de-excitation and excitation operators $\hat{\bm d}^\pm$ in Eq.~\eqref{eq:dipole}. It is clear that the latter operators can be easily generalized in case of more complicated level schemes.

In a similar fashion, we split the electric field operator $\hat{\bm E}(\bm r)=\hat{\bm E}^+(\bm r)+\hat{\bm E}^-(\bm r)$ into two contributions evolving with positive and negative frequencies \cite{mandel:95,walls:95}. The molecule-light coupling is described within the dipole approximation
\begin{equation}\label{eq:hop}
  \hat H_{\rm op}=-\hat{\bm d}\cdot\hat{\bm E}
  \cong -\left(\hat{\bm d}^+\cdot\hat{\bm E}^-+
  \hat{\bm d}^-\cdot\hat{\bm E}^+\right)\,,
\end{equation}
where $\hat{\bm E}(\bm r_m)$ has to be taken at the position $\bm r_m$ of the molecule. We have neglected contributions where both $\hat{\bm d}$ and $\hat{\bm E}$ oscillate with either positive or negative frequencies, which corresponds to the common {\em rotating-wave approximation}\/ \cite{mandel:95,walls:95}. In quantum optics one usually expands the field operators $\hat{\bm E}^\pm$ in the eigenmodes of Maxwell's equation to obtain photon annihilation and creation operators. The first expression in parentheses then describes a molecular de-excitation and the creation of a photon, and the second expression the reversed process. 

Things change considerably if the field operators $\hat{\bm E}$ evolve in a complex dielectric environment where absorption takes place, as is the case for MNPs. Owing to absorption the eigenmodes acquire a complex energy, the imaginary part associated with the damping of the eigenmodes, which spoils the usual quantization procedure for the electromagnetic fields. As we will show next, in linear response one can make use of the {\em fluctuation-dissipation theorem}\/ \cite{kubo:85} to avoid such an eigenmode expansion of $\hat{\bm E}$.

\subsection{Molecule--MNP coupling}

\subsubsection{Green function approach}

Our starting point is provided by the Fermi-golden-rule expression for the molecule decay rate
\begin{equation}\label{eq:goldenrule}
  \gamma=2\pi\, \mu_i\, \left< \hat E_i^+(\bm r_m,\omega)
  \hat E_j^-(\bm r_m,\omega) \right>\,\mu_j
\end{equation}
where we have made use of the Einstein sum convention.\footnote{We use Gauss and atomic units $e=m=\hbar=1$ throughout.} Some details of its derivation are provided in appendix \ref{app:master}. The expression in brackets describes the field fluctuations in thermal equilibrium, to be determined at the molecular transition frequency $\omega$ and position $\bm r_m$. These fluctuations are induced by current noise in the metal. Let $\hat{\bm j}(\bm r)=\hat{\bm j}^+(\bm r)+\hat{\bm j}^-(\bm r)$ denote the current operator in the metal, which we have again split into positive and negative frequency components. If $\bm j$ were a classical current, evolving with frequency $e^{-i\omega t}$, the electric field could be determined from the wave equation \cite{jackson:62}
\begin{equation}\label{eq:wave}
  \nabla\times\nabla\times{\bm E}(\bm r,\omega)-k^2\epsilon(\bm r,\omega)
  {\bm E}(\bm r,\omega)=\frac{4\pi i\omega}{c^2}\,
  {\bm j}(\bm r,\omega)\,,
\end{equation}
where ${\bm E}(\bm r,\omega)$ and ${\bm j}(\bm r,\omega)$ are the Fourier transform of the electric field and current, respectively, $c$ is the vacuum speed of light, $k=\omega/c$ the wavenumber in vacuum, and $\epsilon(\bm r,\omega)$ the space and frequency dependent dielectric function.  We set the magnetic permeability $\mu=1$ throughout. A convenient way to compute the electric field for a given current source is by means of the dyadic Green tensor $\mathbf{G}(\bm r,\bm r',\omega)$ \cite{novotny:06}, to be determined from the wave equation
\begin{equation}\label{eq:green}
  \nabla\times\nabla\times\mathbf{G}(\bm r,\bm r'\omega)-k^2\epsilon(\bm r,\omega)
  \mathbf{G}(\bm r,\bm r',\omega)=4\pi\delta(\bm r-\bm r')\openone\,,
\end{equation}
together with appropriate boundary conditions. The calculation of $\mathbf{G}(\bm r,\bm r',\omega)$ is a common problem in classical electrodynamics, and will be described in more detail further below. For the moment it suffices to assume that the Green function is at our hand. By comparison of Eqs.~\eqref{eq:wave} and \eqref{eq:green} we readily observe
\begin{equation}\label{eq:greenlink}
  \bm E(\bm r,\omega)=\frac{i\omega}{c^2}\int
  \mathbf{G}(\bm r,\bm r'\omega){\bm j}(\bm r',\omega)\,d\bm r'\,.
\end{equation}
Thus, for a linear material response, described by the dielectric function $\epsilon(\bm r,\omega)$, the Green function provides the link between the current source and the electric field. Eq.~\eqref{eq:greenlink} can be also used for the corresponding electric-field and current operators \cite{henry:96}, where $\hat{\bm E}^\pm(\bm r,\omega)$ is induced by $\hat{\bm j}^\pm(\bm r',\omega)$. Note that in the linear response regime positive and negative frequency components do not mix.
We can now use Eq.~\eqref{eq:greenlink} to relate the field fluctuations through 
\begin{eqnarray}
  &&\left< \hat E_i^+(\bm r_m,\omega)\hat E_j^-(\bm r_m,\omega) \right>
    =\frac{\omega^2}{c^4}\int G_{ik}(\bm r_m,\bm r,\omega)\nonumber\\ &&\qquad\times
    \left< \hat j_k^+(\bm r,\omega)\hat j_l^-(\bm r',\omega)\right>
    G_{jl}^*(\bm r_m,\bm r',\omega)\,d\bm rd\bm r'\qquad
\end{eqnarray}
to the current fluctuations in the dielectric. Here, the Green functions describe how the field propagates from the current source to the position of the molecule. In appendix \ref{app:lehmann} we show that for local and isotropic dielectrics the current correlation function can be expressed as $\delta_{kl}\delta(\bm r-\bm r')\,\omega^2\epsilon''(\bm r,\omega)/(2\pi)$, with $\epsilon''(\omega)$ being the imaginary part of the dielectric function. Together with the integral equation \cite{henry:96}
\begin{eqnarray}
  &&\int G_{ik}(\bm r,\bm s,\omega)\,k^2\epsilon''(\bm s,\omega)\,
  G_{jk}^*(\bm r',\bm s,\omega)\,ds\nonumber\\
  &&\qquad\qquad\qquad\qquad\qquad
  =4\pi\,\Im m\left[G_{ij}(\bm r,\bm r',\omega)\right]\,,\quad
\end{eqnarray}
which is obtained by multiplying Eq.~\eqref{eq:green} with $G_{jk}^*$ and subtracting the equation obtained by complex conjugation, we then arrive at the expression for the molecular decay rate
\begin{equation}\label{eq:greendecay}
  \gamma=2k^2\,\bm\mu\cdot\,\Im m\left[\mathbf{G}(\bm r_m,\bm r_m,\omega)\right]
  \,\cdot\bm\mu\,.
\end{equation}
Equation \eqref{eq:greendecay} is our final result. It shows that the decay rate of a molecule in proximity to a MNP is fully determined by the dyadic Green function of classical Maxwell theory. This is a huge simplification because all the dynamics of current fluctuations in the metal are embodied in the dielectric function, which can be obtained from either experiment or first-principles calculation. 

Eq.~\eqref{eq:greendecay} can be brought into an even more transparent form by using the relation $\bm j=-i\omega\bm\mu$ between the current and the molecular dipole. Inserting this expression into Eq.~\eqref{eq:greenlink}, one immediately obtains for Eq.~\eqref{eq:greendecay} the form
\begin{equation}\label{eq:selfinteraction}
  \frac\gamma 2=-\Im m\left[ \bm\mu\cdot\bm E(\bm r_m)\right]\,,
\end{equation}
where $\bm E(\bm r_m)$ is the electric field induced by the dipole $\bm\mu$ itself. Equation~\eqref{eq:selfinteraction} describes a self-interaction, where the dipole polarizes the MNP, and the total electric field of the dipole and of the polarized MNP acts, in turn, back on the dipole. It is the imaginary part of this self interaction which accounts for the decay of the excited molecular state.

\subsubsection{Quasistatic approximation}

In many cases of interest the size of the MNP is much smaller than the light wavelength. It is then possible to employ the so-called {\em quasistatic approximation}.\/ For a given source, the electric field is computed from the scalar potential $\Phi(\bm r)$ according to $\bm E(\bm r)=-\nabla\Phi(\bm r)$. This approach differs from the truly static case in that one uses the frequency-dependent dielectric function $\epsilon(\bm r,\omega)$, at the molecular transition frequency $\omega$, rather than the static limit. The imaginary part of $\epsilon(\bm r,\omega)$ is associated with Ohmic losses of electromagnetic fields inside the metal. The self-interaction expression in the quasistatic limit thus accounts only for nonradiative losses
\begin{equation}\label{eq:gammanr}
  \frac{\gamma_{\rm nr}} 2=-\Im m\left[ \bm\mu\cdot\bm E_{\rm qs}(\bm r_m)\right]\,,
\end{equation}
as indicated by the subscript on $\gamma$. $\bm E_{\rm qs}(\bm r_m)$ is the electric field at the position of the molecule, induced by the molecular dipole, as computed in the quasistatic limit. We have to additionally account for radiative losses. This is done by using the standard expression 
\begin{equation}\label{eq:gammar}
  \gamma_{\rm r}=\gamma_0\,\frac{\left|\bm\mu+\bm\mu_{\rm MNP}\right|^2}{\mu^2}
\end{equation}
for the composite molecule-MNP dipole radiator. Here $\gamma_0=\frac 43n_b\mu^2k^3$ is the free-space decay rate of the molecule, with $n_b$ being the refractive index of the embedding medium. The second term on the right-hand side of Eq.~\eqref{eq:gammar} is the enhancement of the radiative molecular decay in presence of the MNP, where $\bm\mu_{\rm MNP}$ is the MNP dipole induced by the molecule. Below we will show how to compute $\bm E_{\rm qs}(\bm r_m)$ and $\bm\mu_{\rm MNP}$ within a boundary element method approach.

\subsubsection{Scaling}

In the general case there are two length scales which define the problem, namely the size of the nanoparticle and the wavelength of the light. In the quasistatic regime, where the light wavelength is assumed to be much larger than the MNP size, it is a single length scale entering the problem. In this regime one can find a simple scaling law for $\gamma_{\rm nr}$ and $\gamma_{\rm r}$ on the size of the MNP. Suppose that we have computed the two decay rates for a given nanoparticle geometry. We can then scale the whole geometry by a scaling factor $\lambda$, according to $r\to\lambda r$, while keeping all other quantities, such as transition frequency $\omega$ or molecular dipole moment $\bm\mu$, fixed. One can show that such a scaling transforms the scattering rates according to
\begin{equation}\label{eq:scaling}
  \gamma_{\rm nr}(\lambda)=\gamma_{\rm nr}/\lambda^3\,,\quad
  \gamma_{\rm r}(\lambda)=\gamma_{\rm r}\,.
\end{equation}
We will find that Eq.~\eqref{eq:scaling} is an extremely useful expression for estimating the relative importance of non-radiative and radiative decay rates for realistic MNPs.

\subsection{Mie theory}

The electromagnetic response of spherical metallic particles can be solved exactly within Mie theory. This theory provides closed expressions in both the retarded and quasistatic case. Indeed, it is a very rare situation that one can find an exact solution for an experimentally interesting, nontrivial problem. In this work we will be using the Mie theory primarily as a test case for our numerical scheme described below. As there exists a vast amount of literature on Mie theory \cite{vanhulst:81}, in appendix~\ref{app:mie} we only briefly sketch the quasistatic approach and provide the explicit expressions used in our calculations.

\subsection{Boundary element method (BEM)}\label{sec:bem}

In the quasistatic limit the wave equation \eqref{eq:wave} for the electric field transforms to the Poisson equation
\begin{equation}\label{eq:poisson}
  \nabla^2\Phi(\bm r)=-4\pi\rho(\bm r)
\end{equation}
for the scalar potential. Here $\rho(\bm r)$ is the charge distribution of the molecular dipole. The problem can be significantly simplified for a MNP described by a homogeneous dielectric function $\epsilon_m(\omega)$ embedded in a medium with dielectric constant $\epsilon_b$. As long as there is no danger of confusion, we will suppress the frequency dependence of the metal dielectric function and assume that $\epsilon_m$ is evaluated at the molecular transition frequency. We can now use the static Green function $G(\bm r,\bm r')=1/(|\bm r-\bm r'|)$, which is the solution of the Poisson equation
\begin{equation}\label{eq:greenqs}
  \nabla^2 G(\bm r,\bm r')=-4\pi\delta(\bm r-\bm r')
\end{equation}
for a point source located at position $\bm r'$. The solution of Eq.~\eqref{eq:poisson} can then be expressed as \cite{garcia:02,hohenester.prb:05}
\begin{equation}\label{eq:bemindirect}
  \Phi(\bm r)=\int_{\partial\Omega} G(\bm r,\bm s')\sigma(\bm s')\,ds'+
  \Phi_{\rm ext}(\bm r)\,,
\end{equation}
where $\partial\Omega$ denotes the surface of the metallic nanoparticle, $\sigma$ is a surface charge that has to be chosen such that the appropriate boundary conditions are fulfilled, and $\Phi_{\rm ext}$ is the scalar potential for the external charge distribution of the molecular dipole. Indeed, Eq.~\eqref{eq:bemindirect} fulfills the Poisson equation \eqref{eq:poisson} everywhere except at the boundaries. The boundary conditions at $\partial\Omega$ are the continuity of the tangential electric field and $\hat{\bm n}\cdot(\bm D_b-\bm D_m)=0$, where $\hat{\bm n}$ is the unit vector normal to the surface and pointing away from the MNP. $\bm D_b$ is the dielectric displacement at the outer surface, and $\bm D_m$ the displacement at the inner surface. The continuity of the electric field is guaranteed for a potential which is continuous at the boundary, as is the case in Eq.~\eqref{eq:bemindirect}. To implement the second boundary condition, we have to take the surface derivative of $\Phi$ at either side of the boundary. The limit $\bm r\to\bm s$ in Eq.~\eqref{eq:bemindirect} has to be treated with some care.%
\footnote{Let us consider $\lim_{\bm r\to\bm s}\hat{\bm n}\cdot\nabla\int G(\bm r,\bm s')\sigma(\bm s')\,ds'$ for a coordinate system where $\hat{\bm n}=\hat{\bm e}_z$, $\bm r=(0,0,z)$, and $\bm s'=\rho(\cos\phi,\sin\phi,0)$ is given in polar coordinates $\rho$ and $\phi$. We compute the boundary integral within a small circle with radius $R$, within which the surface charge $\sigma$ can be approximated by a constant. The integral then becomes
\begin{displaymath}
  \lim_{z\to 0}\hat{\bm n}\cdot\int\frac{\bm r-\bm s'}%
  {|\bm r-\bm s'|^3}\,ds'\rightarrow\lim_{z\to 0}2\pi z
  \int_0^R\rho d\rho\,(\rho^2+z^2)^{-\frac 32}=2\pi\,.
\end{displaymath}
}
With the abbreviations $F(\bm s,\bm s')=(\hat{\bm n}\cdot\nabla)G(\bm s,\bm s')$ and $\Phi'=(\hat{\bm n}\cdot\nabla)\Phi$, we then get
\begin{equation}\label{eq:bim}
  \lim_{\bm r\to\bm s}\Phi'(\bm r)=
  \int_{\partial\Omega}F(\bm s,\bm s')\sigma(\bm s')\,ds'\pm 2\pi\sigma(\bm s)
  +\Phi_{\rm ext}'(\bm s)\,,\,\,
\end{equation}
where the different signs depend on whether the surface derivative is taken inside or outside the nanoparticle.

At this point it is convenient to change from boundary integrals to boundary elements, suitable for a computational implementation. Within the {\em boundary element method}\/ (BEM) approach, the surface of the nanoparticle is approximated by small surface elements of triangular shape, as shown in Fig.~1 for the example of a spherical particle. In its most simple form, which we will be using in this work, the surface charges are assumed to be located at the center of each triangle $\bm s_i$, and the matrices $G_{ij}$ and $F_{ij}$ connecting different surface elements are obtained from $G(\bm s_i,\bm s_j)$ and $F(\bm s_i,\bm s_j)$.%
\footnote{For the diagonal elements of $G$ we change to polar coordinates $\rho$ and $\phi$, and perform the integration within a triangle where $r(\phi)$ denotes the upper limit of $\rho$ for a given $\phi$. We then obtain
\begin{displaymath}
  \int_0^{2\pi}d\phi\int_0^{r(\phi)}\rho d\rho\,\frac 1\rho=\int_0^{2\pi}r(\phi)\,d\phi\,,
\end{displaymath}
which can be easily calculated numerically. The diagonal elements of $F$ are all zero. 
}
The integral equation \eqref{eq:bim} then reduces to two matrix equations 
\begin{equation}
  \Phi_m'=(\bm F+2\pi\openone)\sigma+\Phi_{\rm ext}'\,,\quad
  \Phi_b'=(\bm F-2\pi\openone)\sigma+\Phi_{\rm ext}'\,,
\end{equation}
where $\bm F$ is the matrix with elements $F(\bm s_i,\bm s_j)$ and the subscript of $\Phi'$ indicates the side on which the surface derivative is taken. We can now use the boundary condition $\epsilon_m\Phi_m'-\epsilon_b\Phi_b=0$ to compute, for a given external charge distribution, the surface charges according to
\begin{equation}\label{eq:bemsigma}
  \sigma=-\bigl\{
  2\pi(\epsilon_m+\epsilon_b)\openone+(\epsilon_m-\epsilon_b)\bm F\bigr\}^{-1}\,
  (\epsilon_m-\epsilon_b)\Phi_{\rm ext}'\,.
\end{equation}
Equation~\eqref{eq:bemsigma} is the main result for our BEM approach. It shows that the surface charge can be computed through numerical inversion of a matrix having the size of the number of triangular surface elements, which is typically of the order of thousands. Once we have computed $\sigma$ for a given inhomogeneity $\Phi_{\rm ext}'$, we can compute the potential and electric fields at any given space point from Eq.~\eqref{eq:bemindirect}, and the induced dipole moment from $\bm\mu_{\rm MNP}=\sum_i\bm s_i\sigma_i$. In our computational approach we describe the molecular dipole by two oppositely charged particles separated by a small distance, and compute the electric field from Eq.~\eqref{eq:bemindirect} by means of a finite difference scheme.

\subsection{Eigenmode expansion}

\subsubsection{Boundary element method approach}

Things can be formulated differently when the dielectric function is approximately of Drude form
\begin{equation}\label{eq:drude}
  \epsilon_m(\omega)=\epsilon_0-\frac{\omega_p^2}{\omega(\omega+i\gamma)}\,.
\end{equation}
Here $\epsilon_0$ is a dielectric constant, accounting for screening of $d$-band electrons in transition metals, $\omega_p$ is the bulk plasma frequency, and $\gamma$ the Landau damping of plasmons. In gold the above Drude form is valid for energies below the threshold value of approximately 2 eV where $d$-band excitations set in. Typical values for Au, which we will be using in this work, are $\epsilon_0=10$, $\omega_p=9$ eV, and $1/\gamma=10$ fs. Instead of the Drude form we can also use a hydrodynamic model where electrons with particle density $n_0$ move freely in a medium with the background dielectric constant $\epsilon_0$. This model is completely equivalent to the framework of the Drude dielectric function. As we will show below, through the hydrodynamic model we can establish a microscopic description for the electron dynamics, which will prove useful for interpreting our results in physical terms and for performing a quantization of surface plasmons.

The energy of a classical electron plasma is the sum of kinetic and electrostatic energy \cite{ritchie:57,barton:79}
\begin{equation}\label{eq:energyplasma}
  H=\mbox{$\frac 12$}\int\left\{n_0(\nabla\Psi)^2+\rho\Phi\right\}d^3r\,.
\end{equation}
Here $\rho(\bm r)$ is the charge density displacement from equilibrium, $\Phi(\bm r)$ is the electrostatic potential induced by $\rho(\bm r)$, and $\Psi(\bm r)$ is the velocity potential, whose derivative gives the velocity density $\bm v=-\nabla\Psi$ \cite{barton:79}. For the surface plasmons of our present concern we consider surface charge distributions $\sigma$ which are nonzero only at the surface of the MNP. As detailed in appendix~\ref{app:spp}, the Hamilton function \eqref{eq:energyplasma} can be rewritten in a boundary element method approach as
\begin{eqnarray}\label{eq:energyspp}
  H&=&\frac 1{2n_0}\Bigl\{\dot\sigma^T\left(2\pi\openone+\tilde{\bm F}\right)^{-1}\bm G\,\dot\sigma
  \nonumber\\&+&
  \omega_p^2\,\sigma^T
  \left[2\pi(\epsilon_0+\epsilon_b)\openone+(\epsilon_0-\epsilon_b)\tilde{\bm F}\right]^{-1}\bm G\,
  \sigma\Bigr\}\,.\qquad
\end{eqnarray}
Here $\sigma$ is the vector of the surface charges within the discretized surface elements, $\omega_p=(4\pi n_0)^{\frac 12}$ is the plasma frequency, and $\tilde{\bm F}$ is the surface derivative of the Green function with respect to the second argument. Equation~\eqref{eq:energyspp} accounts for undamped plasma oscillations. Damping as well as coupling to the molecular dipole has to be included additionally, as we will discuss further below.

Before doing so, we show that Eq.~\eqref{eq:energyspp} allows us to compute eigenmodes for the surface charge oscillations. As the matrices appearing in the Hamilton function are real and symmetric, they can be diagonalized simultaneously. Let $\omega_\lambda^2$ and $u_\lambda$ denote the eigenvectors and eigenvalues of the generalized eigenvalue problem
\begin{eqnarray}\label{eq:eigenvalue}
  &&\omega_\lambda^2\,\left(2\pi\openone+\tilde{\bm F}\right)^{-1}\bm G\,u_\lambda
  \nonumber\\&&\quad=
  \omega_p^2\,
  \left[2\pi(\epsilon_0+\epsilon_b)\openone+(\epsilon_0-\epsilon_b)\tilde{\bm F}\right]^{-1}\bm G\,
  u_\lambda\,.\quad
\end{eqnarray}
The eigenvectors $u_\lambda$ can be chosen real and are orthogonal in the sense $u_\lambda^T\,(2\pi\openone+\tilde{\bm F})^{-1}\bm G\,u_{\lambda'}=  \beta_\lambda\,\delta_{\lambda\lambda'}$, where $\beta_\lambda$ is a real number. We can now expand the surface charge distribution in terms of these eigenfunctions viz. \cite{ritchie:57,barton:79,arista:01}
\begin{equation}\label{eq:sppfield}
  \sigma=\sum_\lambda \left(\frac{2n_0}{\omega_\lambda\beta_\lambda}\right)^{\frac 12}
  u_\lambda(\bm r)\left(a_\lambda^{\phantom*}e^{-i\omega_\lambda t}+
  a_\lambda^*e^{i\omega_\lambda t}\right)\,,
\end{equation}
where $a_\lambda$ are expansion coefficients for the plasmon mode $\lambda$. With Eq.~\eqref{eq:sppfield} the Hamilton function can be brought to the intriguing form $H=\frac 12\sum_\lambda \omega_\lambda(a_\lambda^*a_\lambda^{\phantom*}+a_\lambda^{\phantom*}a_\lambda^*)$. The quantization may now be performed through the substitution $a_\lambda^{\phantom*},\,a_\lambda^*\to \hat a_\lambda^{\phantom\dagger},\,\hat a_\lambda^\dagger$, where $\hat a_\lambda^{\phantom\dagger}$ and $\hat a_\lambda^\dagger$ are the operators for annihilation and creation of a surface plasmon mode $\lambda$, respectively. They obey the usual bosonic commutation relations. With these quantized surface plasmon modes Eq.~\eqref{eq:eigenvalue} can be brought to the form
\begin{equation}\label{eq:hampl}
  \hat H_{\rm pl}=\sum_\lambda\omega_\lambda\left(\hat a_\lambda^\dagger
  \hat a_\lambda^{\phantom\dagger}+\mbox{$\frac 12$}\right)\,.
\end{equation}
This is the Hamilton operator for surface plasmons. In principle it can be used to describe the non-linear properties of metallic nanoparticles, although in this work we shall only be interested in linear response. What still remains to be done is to show how Landau damping can be included, and how to obtain the nonradiative and radiative scattering rates starting from Eq.~\eqref{eq:hampl}.

Let us first consider the coupling between the molecular dipole and the surface plasmons \eqref{eq:sppfield}. Using the expressions given in appendix \ref{app:spp}, we can compute for a given surface mode $u_\lambda$ the scalar potential $\Phi_\lambda$ and its surface derivative $\Phi_\lambda'$ according to
\begin{eqnarray}
  \Phi_\lambda &=& - 4\pi\left[2\pi(\epsilon_0+\epsilon_b)\openone+(\epsilon_0-\epsilon_b)
  \tilde{\bm F}\right]^{-1}\bm G\,u_\lambda\nonumber\\
  \Phi_\lambda' &=& - \bm G^{-1}\left(2\pi\openone-\tilde{\bm F}\right)\Phi_\lambda\,.
\end{eqnarray}
The scalar potential $\tilde\Phi_\lambda$ at the position of the molecular dipole is then obtained by means of Green's second identity [Eq.~\eqref{eq:greensecond}], and the coupling between the molecular dipole and the surface plasmon $u_\lambda$ from the standard expression $g_\lambda=\rho_{\rm dip}\tilde\Phi_\lambda$. In our numerical approach we again describe the charge distribution $\rho_{\rm dip}$ of the molecular dipole by means of two oppositely charged particles separated by a small distance. Within this approach we then obtain for the coupling between the molecule and the surface plasmons the interaction Hamiltonian
\begin{equation}\label{eq:hamplmol}
  H_{\rm pl-mol}=\sum_\lambda \left( 
  g_\lambda^{\phantom*}\,\hat\sigma_+\hat a_\lambda^{\phantom\dagger}+
  g_\lambda^*\,\hat\sigma_-\hat a_\lambda^\dagger\right)\,,
\end{equation}
where we have made use of the rotating-wave approximation. $\hat\sigma_\pm$ are the molecular excitation and de-excitation operators introduced in Eq.~\eqref{eq:dipole}. The first term on the right-hand side of Eq.~\eqref{eq:hamplmol} describes the creation of a surface plasmon through de-excitation of the molecule, and the second term the reversed process of plasmon annihilation and molecule excitation.

\begin{figure}
\centerline{\includegraphics[width=0.95\columnwidth]{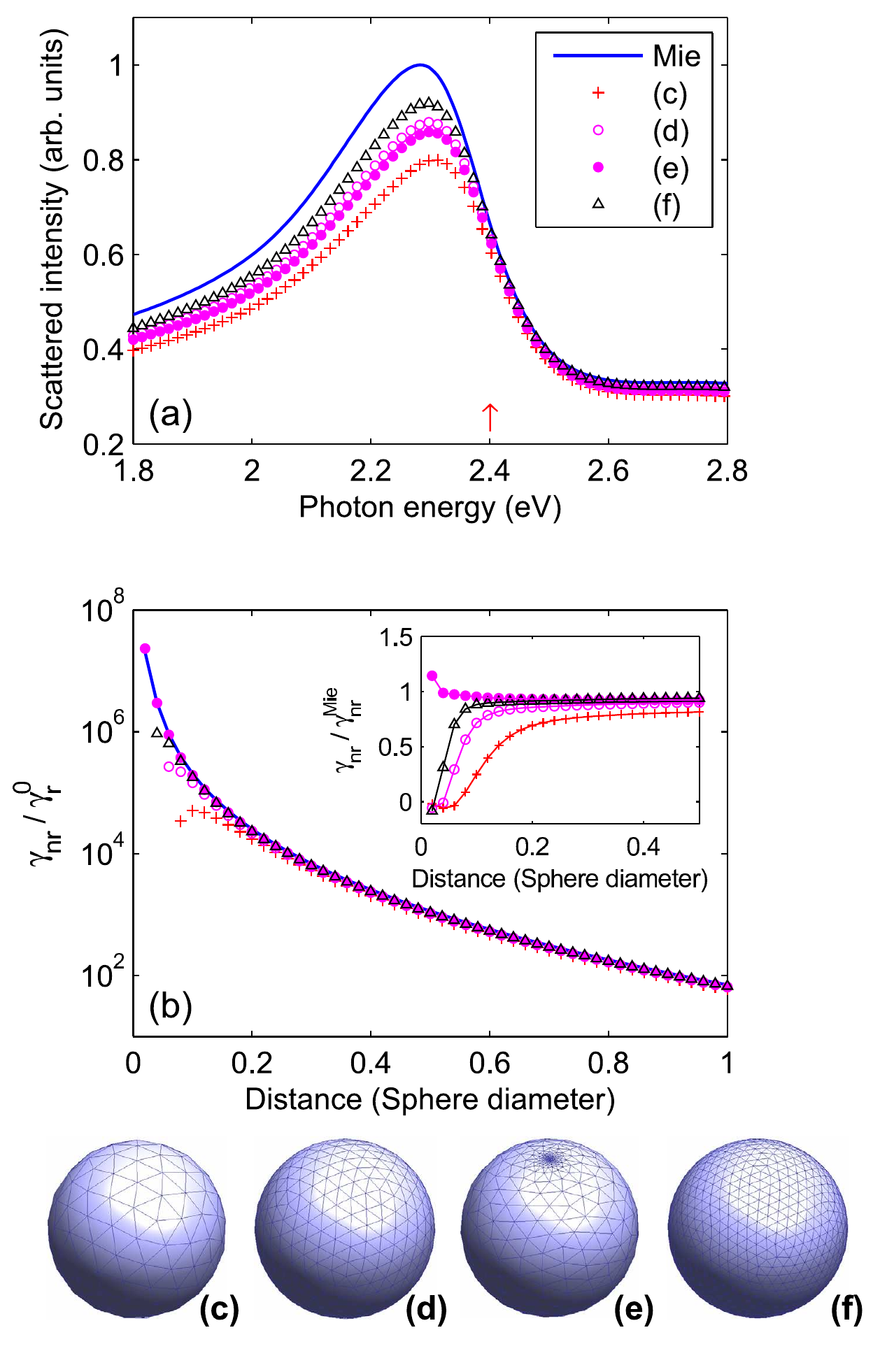}}
\caption{(a) Comparison of BEM and Mie calculations for scattered light intensity of a spherical gold nanoparticle, as computed within the quasistatic approximation and for the dielectric function of Ref.~\cite{johnson:72}. The symbols represent results for different surface discretizations with (c) 144 vertices, (d) 400 vertices, (e) 400 vertices and a finer grid around the north pole, and (f) 900 vertices. (b) Nonradiative decay rate of a molecule in vicinity of the spherical nanoparticle, in units of free-space decay rate $\gamma_r^0$ and for a molecular dipole oriented in radial direction. The transition energy of the molecule is 2.4 eV [see arrow in panel (a)], and the diameter of the nanosphere is 10 nm. The inset reports the ratio of the BEM scattering rate with the exact Mie result.
}\label{fig:mie}
\end{figure}

\subsubsection{Mie theory}

For spherical particles it becomes possible to obtain analytic expressions for the surface plasmon energies $\omega_\lambda$ and coupling constants $g_\lambda$ by means of Mie theory. We here provide for comparison the corresponding results which were first obtained by Ritchie \cite{ritchie:57}. The plasmon energies read
\begin{equation}\label{eq:energymie}
  \omega_l=\omega_p\, \sqrt{\frac l{(l+1)\epsilon_b+l\epsilon_0}}\,,
\end{equation}
where $l$ denotes the usual degrees of spherical harmonics. For the coupling constants we get
\begin{equation}
  g_{lm}=\sqrt{\frac{\omega_l^3}{ln_0}}\,n_ba^{l+\frac 12}\,a_{lm}\,.
\end{equation}
Here $a$ is the radius of the sphere, and $a_{lm}$ are the expansion coefficients of the molecular dipole given in appendix \ref{app:mie}.

\subsubsection{Surface plasmon dynamics}

With the quantization of the surface plasmons we have opened the quantum optics toolbox, which would allow us to investigate strong coupling or other nonlinear effects \cite{walls:95}. In this work, however, we will use the surface plasmon eigenmodes only to analyze the behavior of the radiative and non-radiative decay rates $\gamma_r$ and $\gamma_{nr}$. Pursuing a similar open-system approach as in appendix \ref{app:master}, but for the molecule-plasmon coupling \eqref{eq:hamplmol}, we can express the nonradiative decay rate as
\begin{equation}\label{eq:scattquant}
  \frac{\gamma_{nr}}2=\sum_\lambda \left|g_\lambda\right|^2\int_{-\infty}^0 
  e^{-i\omega t}\left<\hat a_\lambda^{\phantom\dagger}(0)\hat a_\lambda^\dagger(t)\right>\,dt\,.
\end{equation}
Here the exponential accounts for the free propagation of the molecule, and the terms in brackets describe the plasmon correlation function to be evaluated at zero temperature. The latter is of the form $\exp[i(\omega_\lambda-i\frac\gamma 2)t]$, where $\gamma$ is the Landau damping of the plasmons. In a more general approach we could have also introduced damping through a master equation approach with suitable Lindblad operators. We can finally solve the scattering integral of Eq.~\eqref{eq:scattquant} to arrive at
\begin{equation}\label{eq:gammanrquant}
  \frac{\gamma_{nr}}2\cong \sum_\lambda \left|g_\lambda\right|^2 \frac{\frac\gamma 2}%
  {(\omega_\lambda-\omega)^2+\left(\frac\gamma 2\right)^2}\,.
\end{equation}
Again we have neglected the frequency renormalization due to plasmon coupling. Eq.~\eqref{eq:gammanrquant} shows that, through an eigenmode expansion, the nonradiative decay rate can be decomposed into the different contributions for each plasmon mode. Each component is given by the square of the coupling constant together with a Lorentzian whose broadening is given by the Landau damping. Note that in the limit $\gamma\to 0$ the Lorentzian would reduce to $\pi\delta(\omega_\lambda-\omega)$. In addition to Eq.~\eqref{eq:gammanrquant}, the radiative scattering rate can be obtained by computing, within lowest-order perturbation theory, the occupation of the plasmon modes, and multiplying it with the dipole moment of the corresponding mode.

\section{Results}

\subsection{Spherical particles}

\begin{figure}
\centerline{\includegraphics[width=0.85\columnwidth]{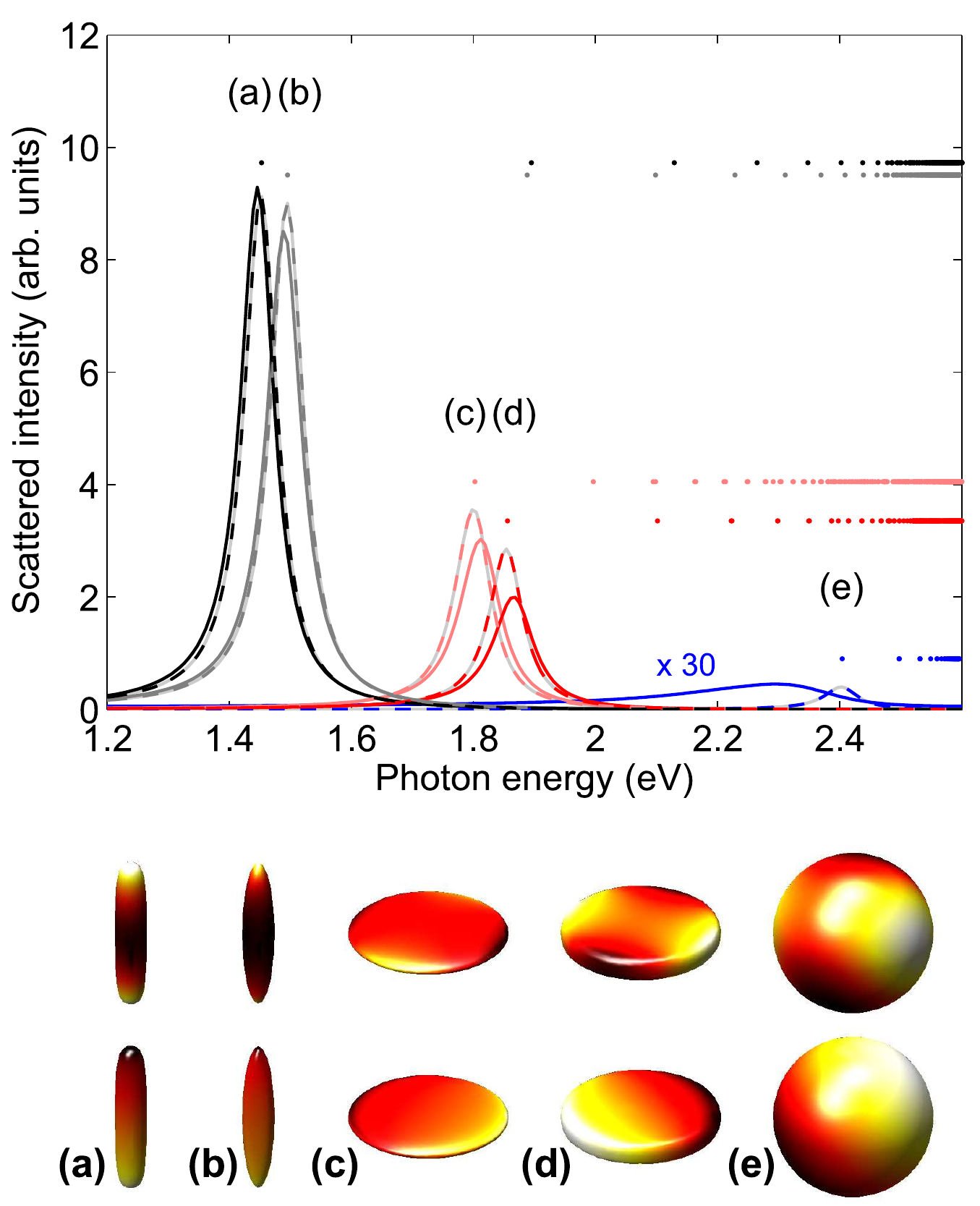}}
\caption{Scattered light intensity for different nanoparticle shapes of (a) rod, (b) cigar, (c) ellipsoid, (d) disk, and (e) sphere. The ratio of height : diameter is $5:1$ for (a,b) and $1:5$ for (c,d). The solid and dashed lines show results obtained for the dielectric function \cite{johnson:72} and the Drude form \eqref{eq:drude}, respectively. Light polarization is chosen along the long axes. The dots above the spectra indicate the positions of the surface plasmon eigenmodes, as obtained from the solutions of Eq.~\eqref{eq:eigenvalue}. The corresponding eigenmodes of lowest energy are shown at the bottom of the figure. For the considered particle shapes only the eigenmodes with lowest energy have a non-zero dipole moment and can couple to light. 
}\label{fig:shape}
\end{figure}

Figure~\ref{fig:mie}(a) reports the scattered light intensity from a spherical gold nanoparticle, as computed within our quasistatic approach, and for the dielectric function of Ref.~\cite{johnson:72}. The solid line is the result of Mie theory, which is summarized in appendix~\ref{app:mie}, and the symbols show results of our boundary element method approach described in Sec.~\ref{sec:bem}. The broad peak at a photon energy of approximately 2.3 eV is associated to the dipole resonance of the nanoparticle. We observe from the figure that the difference between BEM and Mie results reduces with increasing number of surface elements.

Panel (b) shows the nonradiative decay rate for a nanosphere, with 10 nm diameter, and a molecule placed at a given distance away from the particle. We assume that the dipole is oriented in radial direction, and set the molecular transition frequency to 2.4 eV. One observes that with decreasing distance the decay enhancement drastically increases, reaching a value of approximately $10^7$ for the smallest distance of 0.2 nm. The symbols are the results of our BEM calculations. From the inset, which reports the ratio between the BEM and Mie decay rates, one observes that the BEM results give reliable results only down to  molecule-nanoparticle distances comparable to the discretization length. The results for a sphere discretization with 144 vertices [see panel (c)] significantly deviates already at a distance of 2 nm, while the results for sphere discretizations with 400 and 900 vertices [panels (d) and (e)] show nice agreement down to values of about 1 nm. Things improve considerably for the special sphere discretization shown in panel (e). Here we have introduced a finer mesh around the north pole where the molecule approaches the sphere. As evident from the inset of panel (b), the difference between the BEM and Mie calculations is at most ten percent, even for such small distances as 0.2 nm.

\subsection{Other particle shapes}

In the following we consider different particle shapes where the scattering properties cannot be obtained analytically. Figure~\ref{fig:shape} reports optical spectra for different particle shapes, which are shown in the lower panel. In all cases the light polarization is along the long axes of the nanoparticles. The plasmon energy of the dipole mode increases from the quasi one-dimensional rod (a) and cigar (b), over the quasi two-dimensional ellipsoid (c) and disk (d), to the sphere (e). The solid and dashed lines show results as obtained from the dielectric function of Ref.~\cite{johnson:72} and the Drude form \eqref{eq:drude}, respectively. With the exception of the sphere, both results are in nice agreement and thus justify the Drude description for quasi one- and two-dimensional nanoparticles. From the comparison of the results for the rod and cigar, as well as for the ellipsoid and disk, we observe that the detailed shape of the particle has no dramatic impact on the spectra. Finally, the height of the peaks has the approximate ratio $1:4:9$ for the particles of different dimension. Indeed, this is the behavior one would expect for a dipole radiator where the oscillator strength is distributed between (a,b) one, (c,d) two, and (e) three dipole modes, where only one is optically excited.

\begin{figure}
\centerline{\includegraphics[width=0.95\columnwidth]{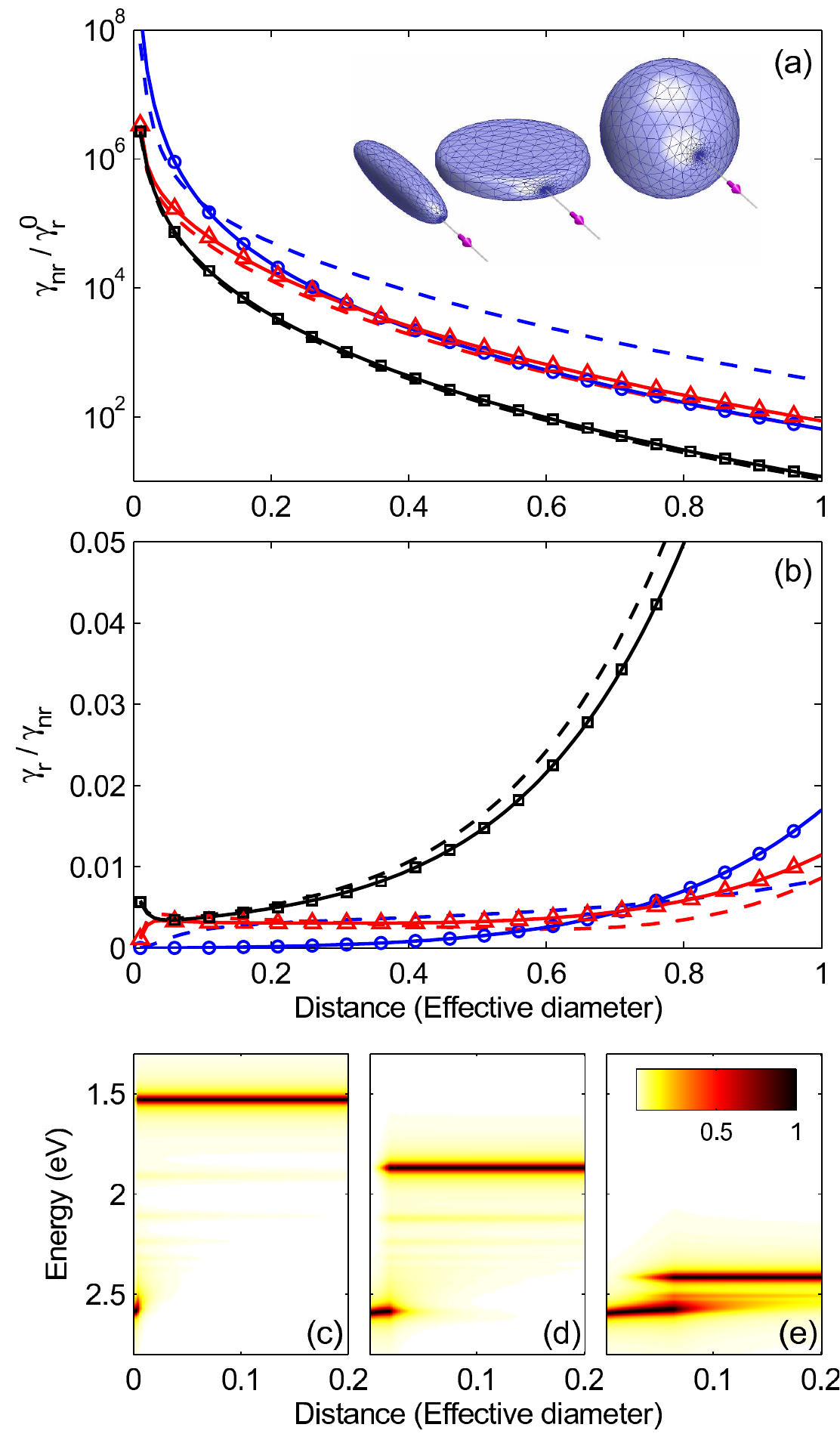}}
\caption{(a) Non-radiative decay rate in units of radiative free-space decay rate $\gamma_r^0$ for different nanoparticle shapes of sphere (circles), disk (triangles), and cigar (squares). The length of the particles along the long axes are 28 nm for the cigar, 16 nm for the disk, and 10 nm for the sphere. These dimensions have been chosen such that all particles have the same volume. The orientation of the molecular dipole and the direction along which the dipole approaches the nanoparticle are indicated in the inset, and the molecular transition energy is assumed to be in resonance with the dipole modes. (b) Quantum yield $\gamma_r/\gamma_{nr}$ for the particles shown in panel (a), and for different molecule-nanoparticle distances measured in units of the sphere diameter of 10 nm. (c) Decomposition of the non-radiative decay rate into eigenmodes, as described in text.
}\label{fig:rate}
\end{figure}

The dots in the upper panel indicate the plasmon eigenenergies $\omega_\lambda$ obtained from the generalized eigenvalue problem \eqref{eq:eigenvalue}. The lowest dipole modes are precisely at the positions of the corresponding peaks in the optical spectra. In the lower panel of the figure we show the eigenmodes $u_\lambda$ for the two surface plasmons of lowest energy. For the particle shapes under consideration only the mode of lowest energy has a finite dipole moment and can thus couple to light. A striking feature of the eigenenergies is that with decreasing energy of the dipole mode the energy separation to the next mode increases. It is of the order of 0.4 eV for the quasi one dimensional particles (a,b), about 0.2 eV for the quasi two dimensional particles (c,d), and decreases by a further factor of two for the sphere. For the sphere we also observe a quasi continuum of surface plasmon modes at energies around 2.5 eV, in agreement with the result \eqref{eq:energymie} of Mie theory. As we will show next, these different energy separations between plasmon states have important consequences for the non-radiative decay rates.

Figure \ref{fig:rate} shows (a) the non-radiative decay rate, and (b) the quantum yield $\gamma_r/\gamma_{nr}$, for the different particle shapes shown in panel (a). In all cases the molecular transition frequency is assumed to be in resonance with the dipole modes. Note that the volume of the particles corresponds to that of a relatively small sphere of 10 nm diameter. By increasing the size of the particles by a factor of, e.g. 2 or 4, $\gamma_{nr}$ would decrease by a factor of 8 and 64, respectively, and the quantum yield increase by the same factor. Thus, for a quasi one- or two-dimensional metallic nanoparticle whose length is around 50 nm, the quantum yield can easily be of the order of several ten percent. 

We can now use Eq.~\eqref{eq:gammanrquant} to decompose the scattering rate into contributions for different plasmon modes $\lambda$. This is what is shown in the lower panels of the figure for (c) the cigar, (d) the disk, and (e) the sphere. Each point in the density plots corresponds to a given molecule-particle distance and plasmon energy, and we have introduced a broadening of the order of $\gamma$ for each plasmon mode. The colors indicate the relative importance of the different plasmon modes. For the cigar-shaped particle, we observe in panel (c) that up to very small distances it is primarily the dipole surface plasmon mode that couples to the molecule. This is because of the large energy separation between the plasmon modes of lowest energy, and the resulting weak coupling to the off-resonant modes. Only for the smallest distances the coupling constants $g_\lambda$ dominate over the detunings $\omega_\lambda-\omega$. Similar behavior is observed in panel (d) for the disk-shaped particle. Things are different in panel (e) showing results for the sphere. Because of the small energy separation between the different surface plasmon modes, already at relatively small distances, say around 0.1 in units of the sphere diameter, the molecule starts to significantly excite the higher-lying plasmon modes. As consequence, the quantum yield dramatically drops at these distances.

\begin{figure}
\centerline{\includegraphics[width=0.95\columnwidth]{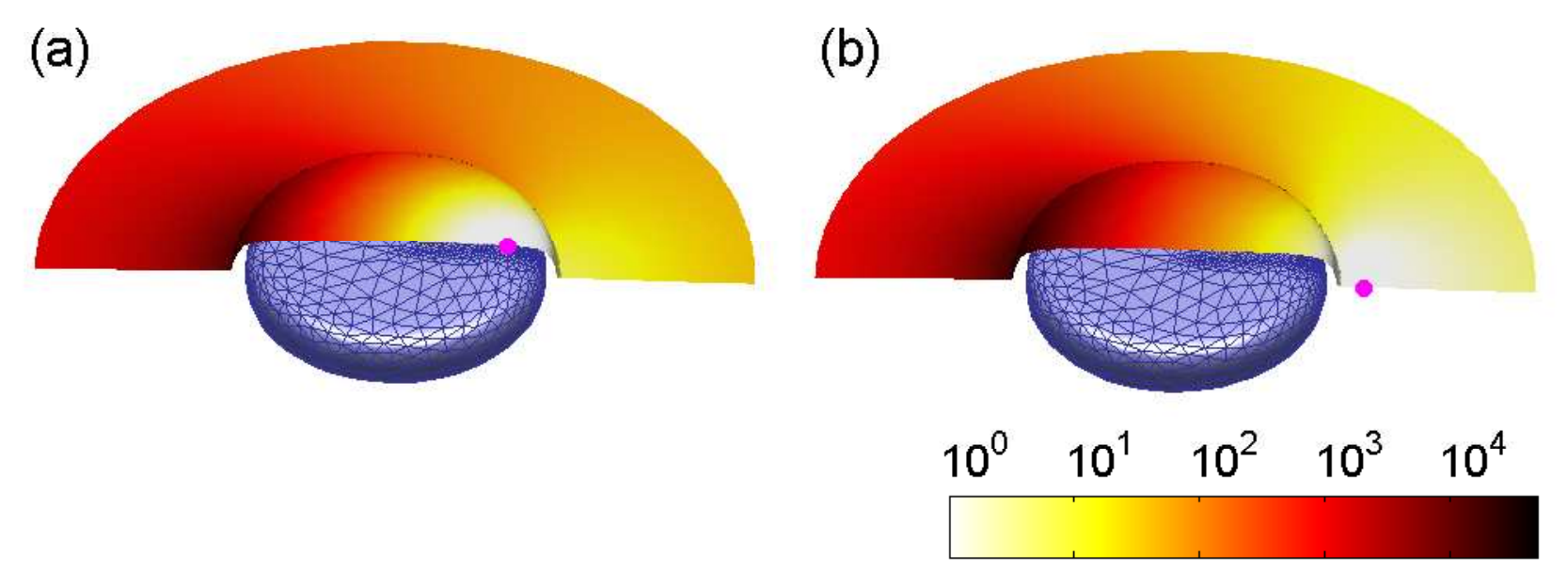}}
\caption{Enhancement of FRET transfer rate for two different donor positions. The transition frequency of the donor molecule is in resonance with the dipole surface plasmon. We perform an average over all dipole orientations of the donor and acceptor molecules. Note that in the quasistatic limit the FRET enhancement does not depend on the particle size.}\label{fig:fret}
\end{figure}

\subsection{F\"orster resonant enery transfer (FRET)}

We conclude this section by briefly discussing F\"orster energy resonance transfer (FRET) in presence of nanoparticles. Here a donor and acceptor molecule exchange energy through dipole-dipole interaction. This process become drastically enhanced if the molecules can benefit from the nearfield enhancement of the metallic nanoparticle \cite{dung:02,govorov:07}. For given donor and acceptor dipole moments $\bm\mu_{\rm don}$ and $\bm\mu_{\rm acc}$, the enhancement of the FRET process can be computed from
\begin{equation}\label{eq:fretenhancement}
  \eta_{\rm FRET}=\frac{\displaystyle\left|\bm\mu_{\rm acc}\cdot
  \bm G(\bm r_{\rm acc},\bm r_{\rm don})\cdot\bm\mu_{\rm don}\right|^2}
  {\displaystyle\left|\bm\mu_{\rm acc}\cdot
  \bm G_0(\bm r_{\rm acc},\bm r_{\rm don})\cdot\bm\mu_{\rm don}\right|^2}\,.
\end{equation}
$\bm r_{\rm don}$ and $\bm r_{\rm acc}$ denote the positions of the donor and acceptor molecule, and $\bm G$ and $\bm G_0$ are the dyadic Green functions in presence and absence of the nanoparticle. In our approach, they are computed in the quasistatic approximation and for the molecular transition frequency, in accordance to the prescription of Sec.~\ref{sec:bem}. Note that the denominator in Eq.~\eqref{eq:fretenhancement} is precisely the free-space dipole-dipole coupling.

Figure \ref{fig:fret} shows the FRET enhancement for a disk-shaped nanoparticle and for a molecular transition frequency equal to the dipole surface plasmon energy. The positions of the donor molecules are indicated in panels (a) and (b). For simplicity, in the plot we assume that the acceptor molecules are placed on a sheet, and we have averaged over all orientation of the donor and acceptor molecules. The FRET transfer becomes enhanced by up to four orders of magnitude, in particular for transfer processes across the nanoparticle. Here, the donor and acceptor molecule exchange their energy via the resonant surface plasmon mode. Note that in order to estimate the FRET efficiency we additionally have to consider the enhancement of the radiative and nonradiative decay processes, as discussed in Ref.~\cite{govorov:07}.


\section{Summary}

In summary, we have developed a general framework, based on a suitable description of quantum electrodynamics in the presence of absorbing media and a general boundary element method approach, for the description of the modified molecular dynamics in presence of metallic nanoparticles. Within the quasistatic limit the decay rate has been separated into radiative and non-radiative decay channels, which both obey a simple scaling behavior. We have discussed different nanoparticle discretizations, and have shown that reliable results can only be obtained for properly defined surface discretizations. From a comparison with Mie theory we have estimated an errorbar of less than ten percent for our numerical results. We have performed an eigenmode expansion for the surface plasmons, and have shown that the higher quantum yield for quasi one- and two-dimensional nanoparticles can be attributed to the large energy splitting of the corresponding eigenmodes. Finally, we have shown that significant enhancements of the F\"orster transfer rates can be expected for donor and acceptor molecules placed in the vicinity of the nanoparticle.

\acknowledgments

We are grateful to Joachim Krenn and Alfred Leitner for most helpful discussions.

\begin{appendix}

\section{Master equation}\label{app:master}

In this appendix we provide some details for the derivation of the Fermi-golden-rule result \eqref{eq:goldenrule}. As we are dealing with an open system, i.e., a molecule interacting with the continuum of photon modes, we have to adopt a density-matrix description \cite{walls:95}. We assume that the molecule is initially in the state $1$. The corresponding density operator is of the form $\hat\rho_0=|1\rangle\langle 1|$. The total density operator $\hat w$ has to additionally account for the degrees of freedom of the electromagnetic environment. Let us make the usual assumption that at time zero the molecule is decoupled from the electromagnetic environment. The total density operator then factorizes $\hat w_0=\hat\rho_0\otimes\hat\rho_R$, where $\hat\rho_R$ is the density operator of the electromagnetic {\em reservoir}\/ which we consider to be in thermal equilibrium. The time evolution of $\hat w(t)$ follows from the Liouville von-Neumann equation $\dot{\hat w}(t)=-i[\hat H_{\rm op}(t),\hat w(t)]$, which we have written in the interaction representation with $H_{\rm op}$ the light-matter coupling of Eq.~\eqref{eq:hop}. As we are only interested in the molecular dynamics, we remove the explicit dependence on the reservoir by tracing out the degrees of freedom of the electromagnetic environment through
\begin{equation}
  \dot{\hat\rho}(t)=\mbox{tr}_R\,\dot{\hat w}(t)=-i\,\mbox{tr}_R
  \left([\hat H_{\rm op}(t),\hat w(t)]\right)
  \,.
\end{equation}
So far we have not achieved too much since the expression on the right-hand side still involves the full density operator $\hat w$. To proceed, we describe the light-matter coupling in lowest order time-dependent perturbation theory, also known as the Born--Markov approximation \cite{walls:95},
\begin{equation}\label{eq:born}
  \dot{\hat\rho}(t)\cong -\mbox{tr}_R\int_{0}^t \left[
  \hat H_{\rm op}(t),\left[\hat H_{\rm op}(t'),\rho(t)\otimes\rho_R\right]\right]\,dt'\,.
\end{equation}
Equation \eqref{eq:born} is correct up to second order in the light-matter coupling $\hat H_{\rm op}$. There is a subtle point regarding the time argument of $\rho(t)$ on the right-hand side, for which we refer the interested reader to the literature \cite{breuer:99,breuer:02}. We next insert the light-matter coupling \eqref{eq:hop} in rotating-wave approximation into Eq.~\eqref{eq:born}, to arrive at the generalized master equation
\begin{eqnarray}\label{eq:master-nonmarkov}
  &&\dot{\hat\rho}(t)\cong -\int_{0}^t dt'\Bigl[ \nonumber\\
  &&\quad\phantom{+}
  \left< \hat E_i^+(t)\hat E_j^-(t')\right>\,
  \hat d_i^-(t)\,\hat d_j^+(t')\,\hat\rho(t)\nonumber\\
  &&\quad+
  \left< \hat E_i^+(t')\hat E_j^-(t)\right>\,
  \hat\rho(t)\,\hat d_i^-(t')\,\hat d_j^+(t)\nonumber\\
  &&\quad-
  \left< \hat E_i^+(t')\hat E_j^-(t)\right>\,
  \hat d_j^+(t)\,\hat\rho(t)\,\hat d_i^-(t')\nonumber\\
  &&\quad-
  \left< \hat E_i^+(t)\hat E_j^-(t')\right>\,
  \hat d_j^+(t')\,\hat\rho(t)\,\hat d_i^-(t)\nonumber\\
  &&\quad+\,\,\,(\pm)\longleftrightarrow(\mp)\Bigr]\,.
\end{eqnarray}
Here we have introduced the shorthand notation $\left<\,.\,\right>= \mbox{tr}_R\left(\hat\rho_R\,.\,\right)$, have exploited the cyclic permutation of the field operators under the trace, and have used that the expectation values $\langle \hat E_i^\pm \hat E_j^\pm\rangle$ vanish in thermal equilibrium. The four terms explicitly given on the right-hand side of Eq.~\eqref{eq:master-nonmarkov} correspond to photon emissions, and those resulting from the exchange of $(\pm)$ with $(\mp)$ to photon absorptions. As in this work we will be dealing with electromagnetic fields in the visible regime and room temperature, we can safely ignore the thermal occupation of the radiation modes, and thus neglect absorption processes of thermal photons throughout.

Let us assume that the dipole operators oscillate with the molecular transition frequency $\omega_0$ according to $\hat{\bm d}^\pm(t)=e^{\mp i\omega_0t}\hat{\bm d}^\pm$. In addition, we introduce the Fourier transform of the field operators
\begin{equation}\label{eq:fourier-field}
  \hat{\bm E}^{\pm}(t)=\int_0^\infty e^{\mp i\omega t}\hat{\bm E}^{\pm}(\omega)
  \,\frac{d\omega}{2\pi}\,.
\end{equation}
We will use the same symbol $\hat{\bm E}$ for the field operators in time and frequency space as long as there is no danger of confusion. For sufficiently large times $t$, where $\omega_0 t\gg 1$ is fulfilled, we can replace the lower limit $0$ in the time integration \eqref{eq:master-nonmarkov} by $-\infty$. This replacement is called the {\em adiabatic approximation}\/ \cite{walls:95,breuer:02} and is usually considered to be valid under broad conditions. The time integrations then reduce to integrals over exponentials, which can be solved, with the relative time $\tau=t-t'$, according to
\begin{equation}\label{eq:adiabatic}
  \int_{-\infty}^0 e^{-i(\omega_0-\omega+i0^+)\tau}d\tau=
  \frac i{\omega_0-\omega+i0^+}\cong \pi\delta(\omega-\omega_0)\,.
\end{equation}
The infinitesimally small and positive quantity $0^+$ has been introduced to provide a damping of the exponential at early times. Its appearance has nothing to do with a real damping process. Rather it allows to perform the adiabatic limit in the integration {\em prior}\/ to the thermodynamic limit in the expectation value $\left< \,.\,\right>$ \cite{fick:90}. Finally, in the last expression on the right hand side of Eq.~\eqref{eq:adiabatic} we have only considered Dirac's delta function, which accounts for the energy conservation in the scattering processes, and have neglected the Cauchy principal value. The latter contribution is responsible for an energy shift of the molecular transition, usually known as the {\em Lamb shift}\/ \cite{walls:95}, which could be incorporated into a renormalized $\omega_0$.
In thermal equilibrium, the expectation value of the field operators $\langle \hat E_i^+(\omega)\hat E_j^-(\omega')\rangle=2\pi\delta(\omega-\omega')\langle \hat E_i^+(\omega)\hat E_j^-(\omega)\rangle$ is diagonal in frequency space. Putting together all the results and changing back from the interaction to the Schr\"odinger representation, then brings us to the master equation
\begin{eqnarray}\label{eq:master}
  \dot{\hat\rho}&\cong&-i\left[\hat H_0,\hat\rho\right]-
  \pi\left< \hat E_i^+(\omega_0)\hat E_j^-(\omega_0)\right>\nonumber\\
  &\times&
  \left(\hat d_i^-\hat d_j^+\,\hat\rho+\hat\rho\,\hat d_i^-\hat d_j^+
  -2\hat d_j^+\,\hat\rho\,\hat d_i^-\right)\,.\,\,
\end{eqnarray}
The first term on the right hand side accounts for the free propagation of the molecule, and the second term for scattering processes due to the coupling to the electromagnetic fields. The terms in parentheses are reminiscent of the Lindblad master equation. Finally, using the explicit form \eqref{eq:dipole} for the dipole operators and taking the matrix elements $\rho_{11}=\langle 1|\hat\rho|1\rangle$ of the operator equation \eqref{eq:master}, we find that the upper-state population of the molecule decays according to $\dot\rho_{11}=-\gamma\,\rho_{11}$, with the decay rate $\gamma$ given in Eq.~\eqref{eq:goldenrule}.

\section{Current correlation function}\label{app:lehmann}

In this appendix we provide some details of how to relate the current correlation function $\langle \hat j_k^+(\bm r,\omega)\hat j_l^-(\bm r',\omega)\rangle$ to the imaginary part of the dielectric function of the metal. We will only consider local and isotropic media, such that the expression given above reduces to $\delta_{kl}\delta(\bm r-\bm r')\langle \hat j^+(\omega)\hat j^-(\omega) \rangle$. The calculation of such correlation functions is a common problem in solid state physics \cite{fetter:71,mahan:81}.

We start with a general result of linear response theory
\begin{equation}
  \epsilon''(\omega)=\frac{4\pi}{\omega^2}\Im m
  \int_0^\infty e^{i\omega t} \left<[\hat j^+(t),\hat j^-(0)]\right>\,dt\,,
\end{equation}
which relates the imaginary part of the dielectric function to the retarded current correlation function \cite{kubo:85}. A common link between the ordered and retarded correlation function is provided by the spectral function $\rho(t)=\left<[\hat j(t),\hat j^\dagger(0)]\right>$. To compute $\rho(t)$ we would need a microscopic model for the electron dynamics in the metal, which is certainly beyond the scope of our work. However, in order to obtain, in thermal equilibrium, the relation between the different correlation functions it suffices to assume that the states $|m\rangle$ and energies $E_m$ of the full many-body hamiltonian $\hat H$ could be determined {\em in principle}.\/ In fact, we never have to compute these states explicitly. Inserting in $\langle\,.\,\rangle=Z^{-1} \, \mbox{tr}(e^{-\beta\hat H}\,.\,)$ a complete set of states, where $\beta=1/(k_BT)$ is the inverse temperature and $Z$ the partition function, we then obtain for the Fourier transform of the spectral function \cite{fetter:71,mahan:81} 
\begin{eqnarray}\label{eq:spectraldensity}
  \rho(\omega)&=&\left(1-e^{-\beta\omega}\right)Z^{-1}\sum_{m,n}e^{-\beta E_m}
  \left|\langle m|\hat j^+|n\rangle\right|^2\nonumber\\
  &&\qquad\times 2\pi\delta(\omega-[E_n-E_m])\,.
\end{eqnarray}
From Eq.~\eqref{eq:spectraldensity} one then obtains $\langle\hat j^+(\omega)\hat j^-(\omega)\rangle=\rho(\omega)/(1-e^{-\beta\omega})$. Similarly, upon insertion of the many-body states into the retarded correlation function we get $\rho(\omega)=\omega^2\,\epsilon''(\omega)/(2\pi)$. Putting together all results we arrive at our final expression
\begin{equation}
  \left<\hat j^+(\omega)\hat j^-(\omega)\right>=
  \left[\bar n_{\rm th}(\omega)+1\right]\,\frac{\omega^2\,\epsilon''(\omega)}{2\pi},,
\end{equation}
with $\bar n_{\rm th}(\omega)=1/(e^{\beta\omega}-1)$ the Bose-Einstein distribution function. For the problem of our present concern we can safely neglect the thermal occupation $\bar n_{\rm th}$ of photons.

\section{Mie theory in the quasistatic limit}\label{app:mie}

We consider the situation where a single molecule with dipole moment $\bm\mu$ is located at some distance away from a spherical nanoparticle with radius $a$. We expand the potential inside and outside the particle in terms of spherical harmonics $Y_{lm}$ \cite{jackson:62}
\begin{eqnarray}
  \Phi_{\rm in} &=& \sum_{l=0}^{\infty}\sum_{m=-l}^l\frac{4\pi}{2l+1}\, b_{lm}r^lY_{lm}(\theta,\phi)\nonumber\\
  \Phi_{\rm out} &=& \sum_{l=0}^{\infty}\sum_{m=-l}^l\frac{4\pi}{2l+1}\, c_{lm}\frac{Y_{lm}(\theta,\phi)}{r^{l+1}}+\Phi_{\rm dip}\,.
\end{eqnarray}
Here $b_{lm}$ and $c_{lm}$ are expansion coefficients, to be determined below, $r$ is the distance from the center of the sphere, and $\Phi_{\rm dip}$ is the potential of the molecular dipole. $\Phi_{\rm dip}$ can be expanded similar to the first expression for $\Phi_{\rm out}$, but with different expansion coefficients $a_{lm}$. They read \cite{gersten:91}
\begin{eqnarray}\label{eq:miedipole}
  a_{lm}&=&-\frac 1{r^{l+2}}\Bigl\{(l+1)(\hat{\bm n}\cdot\bm\mu)Y_{lm}^*(\theta,\phi)\nonumber\\
  &&\qquad\quad+i\sqrt{l(l+1)}(\hat{\bm n}\times\bm\mu)\cdot\bm X_{lm}^*(\theta,\phi)\Bigr\}\,.\qquad
\end{eqnarray}
Here $\hat{\bm n}$ is the unit vector from the center of the sphere to the dipole located at the position characterized by $r$, $\theta$, $\phi$, and $\bm X_{lm}(\theta,\phi)$ are the vector spherical harmonics \cite{jackson:62}.
We can now use the boundary conditions of Maxwell's theory to relate the different coefficients viz.
\begin{equation}
  c_{lm}=\frac{(1-\epsilon_r)la^{2l+1}}{(1+\epsilon_r)l+1}\,a_{lm}\,,
\end{equation}
with $\epsilon_r=\epsilon_m/\epsilon_b$ being the ratio between the dielectric functions of the MNP and the background material within which the particle is embedded. The electric field, needed for the calculation of the non-radiative decay rate \eqref{eq:gammanr}, becomes
\begin{eqnarray}
  \bm E_{\rm out}&=&\sum_{l=0}^{\infty}\sum_{m=-l}^l\frac{4\pi}{2l+1}\,\frac{c_{lm}}{r^{l+1}}
  \bigl\{(l+1)\hat{\bm n}\,Y_{lm}(\theta,\phi)\nonumber\\
  &&\qquad+
  i\sqrt{l(l+1)}\,\hat{\bm n}\times\bm X_{lm}^*(\theta,\phi)\bigr\}+\bm E_{\rm dip}\,.\qquad
\end{eqnarray}
For the dipole moment of the MNP, induced by the molecule, we get \cite{jackson:62}
\begin{equation}
  \bm\mu_{\rm MNP}=\sqrt{\frac{4\pi}3}\left(c_{10}\hat{\bm e}_z+c_{11}\hat{\bm e}_+
  -c_{11}^*\hat{\bm e}_-\right)\,.
\end{equation}
Here $\hat{\bm e}_z$ is the unit vector along $z$, and $\hat{\bm e}_\pm=(\hat{\bm e}_x\pm i\hat{\bm e}_y)/\sqrt{2}$.

\section{Hydrodynamic model}\label{app:spp}

In this appendix we show how to obtain Eq.~\eqref{eq:energyspp} starting from expression \eqref{eq:energyplasma} for the energy of a classical plasma. Let us consider first the potential energy term. Our starting point is provided by Green's second theorem
\begin{equation}\label{eq:greensecond}
  4\pi\Phi(\bm r)=
  \int_{\partial\Omega}\Bigl( 
  G(\bm r,\bm s')\frac{\partial\Phi(\bm s')}{\partial\hat n'}-
  \frac{\partial G(\bm r,\bm s')}{\partial\hat n'}\Phi(\bm s')\Bigr)\,ds'\,,\quad
\end{equation}
where the partial derivatives denote the surface derivatives $\hat{\bm n}\cdot\nabla$ with respect to $\bm s'$. Similarly to the discussion in Sec.~\ref{sec:bem}, we can perform in Eq.~\eqref{eq:greensecond} the limit $\bm r\to\bm s$ to obtain a relation between $\Phi$ and its surface derivative. Within our boundary element method approach we obtain $(2\pi\openone\pm\tilde{\bm F})\Phi=\pm\bm G\Phi'$ for the limit taken in- or outside the metallic nanoparticle. $\tilde{\bm F}$ denotes the surface derivative of $G$ with respect to the second argument. To obtain a relation between $\Phi$ and the surface charge $\sigma$, we make use of the boundary condition $\hat{\bm n}\cdot(\bm D_b-\bm D_m)=4\pi\sigma$. Together with Eq.~\eqref{eq:greensecond} we then obtain
\begin{equation}\label{eq:phi-sigma}
  \Phi=4\pi\bigl\{2\pi(\epsilon_0+\epsilon_b)\openone+(\epsilon_0-\epsilon_b)\tilde{\bm F}
  \bigr\}^{-1}\bm G\,\sigma\,.
\end{equation}
As for the kinetic energy of the classical plasma, we use the continuity equation
\begin{equation}\label{eq:continuity}
  \partial_t n=-n_0\nabla\cdot\bm v=n_0\nabla^2\Psi\,,
\end{equation}
to relate the density displacement $n$ and the velocity potential $\Psi$. Integration of the continuity equation \eqref{eq:continuity} over a small cylinder $\Omega$ (height $h\to 0$ and base $\delta S$) containing a small surface element, then gives for the right-hand side of Eq.~\eqref{eq:continuity}
\begin{equation}\label{eq:intcontinuity}
  \int_\Omega n_0\nabla^2\Psi\,d^3r=\int_{\partial\Omega} n_0\,\hat{\bm n}\cdot\nabla\Psi\,ds
  \cong n_0\,\frac{\partial\Psi}{\partial\hat n}\,\delta S\,.
\end{equation}
Here, $\frac{\partial\Psi}{\partial\hat n}=\hat{\bm n}\cdot\nabla\Psi$ denotes the surface derivative of the velocity potential. Together with the left-hand side of Eq.~\eqref{eq:continuity} we find the link between $\Psi$ and $\sigma$,
\begin{equation}\label{eq:sigma-psi}
  n_0\,\frac{\partial\Psi}{\partial\hat n}=\dot\sigma\,.
\end{equation}
Using Green's second identity, we can again relate within our boundary element method approach the surface derivative of $\Psi$ to the velocity potential viz. $\Psi=(2\pi\openone+\tilde{\bm F})^{-1}\bm G\,\Psi'$. We can thus express the kinetic energy 
\begin{equation}\label{eq:energypot}
  \mbox{$\frac 12$}n_0\int_{\partial\Omega}\Psi\frac{\partial\Psi}{\partial\hat n}\,ds'
  \longrightarrow\mbox{$\frac 12$}n_0\,\dot\sigma^T(2\pi\openone+\tilde{\bm F})^{-1}
  \bm G\,\dot\sigma\,.
\end{equation}
through the time derivative of the surface charge distribution. Putting finally together the contributions for the kinetic and potential energies, we arrive at expression \eqref{eq:energyspp} for the energy of a classical plasma in terms of the surface charge $\sigma$.

\end{appendix}


\end{document}